\documentclass[12pt,a4paper,leqno]{article}
\usepackage{amsmath}
\title{Discussion on exp-function  method and  
modified method of simplest equation}
\author{Zlatinka I. Dimitrova}
\date{}
\begin{document}
\maketitle
\begin{abstract}
We discuss the relation between the modified method of simplest equation and the
exp-function method. First on the basis of our experience from the application of the method of simplest equation we generalize the exp-function ansatz.  Then we apply the ansatz for obtaining exact 
solutions for members of a class of nonlinear PDEs which contains as particular cases several
nonlinear PDEs that model the propagation of water waves.
\end{abstract}
\begin{flushleft}
{\bf Key words:} 
\end{flushleft}
nonlinear partial differential equations,  method of simplest equation, exact traveling-wave 
solutions, exp-function method \vskip1cm
\section{Nonlinear PDEs and method of simplest equation}
Nonlinear models are much used in various branches of science [$^{1,2,3}$]. 
Often such models contain nonlinear PDEs and because of this the interest
in obtaining exact analytical solutions of  nonlinear PDEs increses steadily.  
Such exact solutions often describe important classes of waves and 
processes in the investigated systems. In addition the exact solutions can be 
useful as initial conditions in the  process of obtaining of numerical 
solutions or as test soluitions for  computer programs for obtaining numerical 
solutions of the studied nonlinear PDEs. 
\par
Because of all above the  nonlinear PDEs are widely applied in the theory
of solitons [$^{4,5}$], hydrodynamics and theory of turbulence [$^{6}$] - [$^{10}$],
theory of dynamical systems, chaos [$^{11}$] - [$^{14}$ ], etc.
Sophisticated methods for obtaining exact solutions of nonlinear PDEs such as  
the inverse scattering transform or the method of Hirota [$^{15}$] allow 
obtaining of soliton solutions of some equations. In the last several 
years effective approaches for obtaining exact special solutions of complicated
nonlinear nonintegrable 
PDEs have been developed too [$^{16}$] - [$^{18}$]. These approaches leaded to 
 exact solutions of many equations such as the Kuramoto-Shivasinsky 
equation [$^{19}$] or equations, connected to the models of migration of
populations[$^{20, 21, 22}$]. The discussion below will be devoted to 
the modified method of simplest equation: a version of the method of simplest 
equation for  obtaining exact solutions of nonlinear PDEs and on the relation 
of this method  to another popular method: the exp-function  method. 
\par
A brief description of the method of simplest equation is as follows[$^{23}$].
Let us have a partial differential equation and let by means of an
appropriate ansatz this equation be reduced to the nonlinear
ordinary differential equation
\begin{equation}\label{gen_eq}
P \left( F(\xi),\frac{d F}{d \xi},\frac{d^{2} F}{d \xi^{2}},\dots \right) = 0.
\end{equation}
For large class of equations from the kind (\ref{gen_eq}) exact solution 
can be constructed as finite series
\begin{equation}\label{solution}
F(\xi) = \sum_{\mu=0}^{\nu} p_{\mu} [\Phi (\xi)]^{\mu},
\end{equation}
where  $\nu>0$, $\mu$, $p_{\mu}$ are  parameters and $\Phi (\xi)$ is a solution of 
some ordinary differential equation
referred to as the simplest equation. The simplest equation is of lower 
order than (\ref{gen_eq}) and we know the general solution of
the simplest equation or we know at least exact analytical particular
solution(s) of the simplest equation. 
\par
The application of the modified method of simplest equation is 
as follows. First by means of an appropriate ansatz (for an example the 
traveling-wave ansatz) the solved 
class of nonlinear PDEs is reduced to a class of nonlinear ODEs of the 
kind (\ref{gen_eq}). In the method of simplest equation the resulting ODEs
are treated as in the first step of the test for Painleve property: the
corresponding equation is  subject of leading order analysis that leads to 
determination of $\nu$ from Eq.(\ref{solution}). In the modified method
of simplest equation one uses the equivalent procedure of obtaining and
solving a balance equation as follows. First
the finite-series solution (\ref{solution}) is substituted in 
(\ref{gen_eq}) and as a  result a polynomial of $\Phi(\xi)$ is obtained. 
Eq. (\ref{solution}) is a solution of 
(\ref{gen_eq}) if all coefficients of the obtained polynomial of $\Phi(\xi)$ 
are equal to $0$.
Then by means of a balance equation one ensures that there are at least 
two terms in the coefficient of the largest power of $\Phi(\xi)$. The balance 
equation gives a relationship between the parameters of the solved class of 
equations and the parameters of the solution.
The application of the balance equation and setting the coefficients of the 
polynomial of $\Phi(\xi)$ to $0$ leads to a system of nonlinear relationships 
among the parameters of the solution and the parameters of the solved class 
of equations.
Each solution of the obtained system of nonlinear algebraic equations leads to a solution 
of  a nonlinear PDE from the investigated class of  nonlinear PDEs.
\section{The exp-function method: one possible generalization and application}
Let us now consider  the exp-function method.
The standard exp-function ansatz for a solution of a nonlinear 
partial differential equation is $[^{24}]$
\begin{equation}\label{e1}
u(x,t) = \cfrac{\sum_{i=0}^m a_i \exp(i \xi)}{\sum_{j=0}^n b_j \exp(j \xi)}, 
\hskip.25 cm \xi = kx + wt + \delta.
\end{equation}
\par
The authors of the exp-function method do not define the class of 
equations for which an exact solution can be obtained by this method.
In our opinion the method can be applied to some nonintegrable
partial differential equations with polynomial nonlinearity.
The  ansatz (\ref{e1})  can be generalized on the basis of the following observation. In one of 
the variants of the modified method of simplest equation the one-wave solution of the studied
nonlinear partial differential equation is searched by the ansatz
\begin{equation}\label{e4}
u(\xi) = \sum_{l=0}^L A_l [F(\xi)]^{B_l}.
\end{equation}
where $F(\xi)$ is a solution of the simplest equation, $A_l$ and $B_l$ are parameters (if $B_l =l$ Eq. (\ref{e4}) is a 
polynomial ansatz but $B_l$ can be non-integer number too), and $\xi = x - vt + \xi_0$ where $v$ is the velocity of the 
wave and $\xi_0$ is a parameter. When the equation of Bernoulli
\begin{equation}\label{e6}
\frac{d F}{d \xi} = a F(\xi) + b[F(\xi)]^M, \hskip.25cm M=2,3,\dots,
\end{equation}
is used as a simplest equation its solutions are
\begin{eqnarray}\label{e7}
F(\xi) = \left \{ \frac{a \exp[a (M-1) (\xi + \xi_0)]}{1 - b \exp [a (M-1) 
(\xi + \xi_0)]} \right \}^{\frac{1}{M-1}},
{\rm case \ a>0, b<0}, \nonumber \\
F(\xi) = \left \{ - \frac{a \exp[a (M-1) (\xi + \xi_0)]}{1 + b \exp [a (M-1) (\xi + \xi_0)]} \right \}^{\frac{1}{M-1}}
, {\rm case \ a<0, b>0}.
\end{eqnarray}
We observe that the term in $\{ \dots \}$  from Eqs. (\ref{e7})  can be easily obtained from Eq. (\ref{e1}) 
when $n=m=1$. 
\par
On the basis of all above the following simple generalization of the exp-function ansatz is 
obtained: One searches for exact solution of the studied nonlinear PDE on 
the basis of the ansatz
\begin{equation}\label{e9}
u(x,t) = \sum_{l=0}^L A_l \left[ \cfrac{\sum_{i=0}^m a_i \exp(i \xi)}{\sum_{j=0}^n b_j \exp(j \xi)} \right]^{B_l}, 
\hskip.25 cm \xi = kx + wt + \delta
\end{equation}
\par
Let us now apply the above ansatz to the equation
\begin{equation}\label{a1}
\frac{\partial u}{\partial t} - \frac{\partial^3 u}{\partial x^2 \partial t} + \frac{\partial}{\partial x} \left(\sum_{h=1}^H \alpha_h u^h\right)
- \nu \frac{\partial u}{\partial x} \frac{\partial^2 u}{\partial x^2} - \gamma u \frac{\partial^3 u}{\partial x^3} =0
\end{equation}
Particular cases of Eq. (\ref{a1}) are for an example: (I)
The Camassa - Holm equation
$\frac{\partial u}{\partial t} - \frac{\partial^3 u}{\partial x^2 \partial t} + 3 u \frac{\partial u}{\partial x} -  
2 \frac{\partial u}{\partial x} \frac{\partial^2 u}{\partial x^2} - u \frac{\partial^3 u}{\partial x^3}  =0
$,
that describes the propagation of shallow water waves over a flat bottom; (II)
The Degasperis - Processi equation
$
\frac{\partial u}{\partial t} - \frac{\partial^3 u}{\partial x^2 \partial t} + 4 u \frac{\partial u}{\partial x} -  
3 \frac{\partial u}{\partial x} \frac{\partial^2 u}{\partial x^2} - u \frac{\partial^3 u}{\partial x^3}  =0
$
that is also connected to the dynamics of the nonlinear shallow water waves;
(III) The Fornberg - Whitham equation
$
\frac{\partial u}{\partial t} - \frac{\partial^3 u}{\partial x^2 \partial t} + \frac{\partial u}{\partial x} + 
u \frac{\partial u}{\partial x} -  3 \frac{\partial u}{\partial x} \frac{\partial^2 u}{\partial x^2} - u \frac{\partial^3 u}{\partial x^3}  =0
$
that is used as a model for investigation of the wave - breaking.
We shall search for solutions of the kind (\ref{e9}) with $B_l=l B$ where $B$ is a parameter. We shall discuss the simplest possible case $n=m=1$.
For this case first we shall write a balance equation for the maximum powers in the numerators of the terms from Eq. (\ref{e9}).
There are 5 terms in Eq. (\ref{e9}). Each term has several powers of $\exp(\xi)$ in its numerator. From these several powers one is
the maximum power for the corresponding term of Eq. (\ref{e9}). As the terms in Eq. (\ref{e9}) are 5 we have 5 maximum powers.
We impose the balance equation constraint: The two largest of these 5 maximum powers must be equal (otherwise some parameters
of the solution  and eventually some parameters of the equation will be $0$ which in the most of the cases is undesirable). The equality
of the two largest powers leads to the balance equation.
\par
The balance equation constraint leads to the following two possibilities for balance equations
\begin{equation}\label{a6}
2 L B + 3 = 2 L B +3 \hskip.5cm {\rm when} \hskip.5cm H < 2 + \frac{2}{L B}
\end{equation}
and
\begin{equation}\label{a7}
H = 2 + \frac{2}{L B}
\end{equation}
We note that $L$ and $H$ are integers and because of this the parameter $B$ must have appropriate (even non-integer) values.
Let us illustrate this point further by means of two small tables for the two possibilities for balance equations.
\begin{table}[h]
\begin{center}
\begin{tabular}{|c|c|c|c|}
\hline
  &   &                         \\
L & B & $2 L B + 3$ & $H < 2 + \frac{2}{L B}$ \\
  &   &             &            \\
\hline \hline
1 & 1 & 5 & $H<4$ \\
\hline
2 & 1 & 7 & $H<3$ \\
\hline
1 & 1/2 & 4 & $H<6$ \\
\hline
2 & 1/2 & 5 & $H<4$ \\
\hline
\dots & \dots & \dots & \dots \\
\hline
\end{tabular}
\end{center}
\caption {Several of possible values of the parameters for the case when Eq. (\ref{a6}) is balance equation.}
\end{table}

\begin{table}[h]
\begin{center}
\begin{tabular}{|c|c|c|}
\hline
  &   &                         \\
L & B & $H = 2 + \frac{2}{L B}$ \\
  &   &                         \\
\hline \hline
1 & 1 & 4 \\
\hline
2 & 1 & 3 \\
\hline
1 & 1/2 & 6 \\
\hline
2 & 1/2 & 4 \\
\hline
3 & 1/3 & 4 \\
\hline
\dots & \dots & \dots \\
\hline
\end{tabular}
\end{center}
\caption {Several of possible values of the parameters for the case when Eq. (\ref{a7}) is balance equation.}
\end{table}
Let us now consider one examples. Let Eq. (\ref{a7}) be the balance equation and in 
addition $B=1$ and $L=1$. Then $H=4$. Thus the equation we shall solve is
\begin{equation}\label{k1}
\frac{\partial u}{\partial t} - \frac{\partial^3 u}{\partial x^2 \partial t} + \frac{\partial}{\partial x} \left(\sum_{h=1}^4 \alpha_h u^h\right)
- \nu \frac{\partial u}{\partial x} \frac{\partial^2 u}{\partial x^2} - \gamma u \frac{\partial^3 u}{\partial x^3} =0
\end{equation}
The ansatz (\ref{e9}) becomes
\begin{equation}\label{k2}
u(x,t) = \sum_{l=0}^1 A_l \left[ \cfrac{\sum_{i=0}^m a_i \exp(i \xi)}{\sum_{j=0}^n b_j \exp(j \xi)} \right]^{ l}, 
\hskip.25 cm \xi = kx + wt + \delta
\end{equation}
The substitution of Eq. (\ref{k2}) in Eq. (\ref{k1}) leads to
a system of 5 nonlinear algebraic relationships among the parameters of
the solution and the parameters of the equation. One solution of this
system is 
\begin{itemize}
\item $a>0$, $b<0$: 
$a_0 =0$; $a_1 =a$; $b_0 =1$, $b_1=-b$
\item $a<0$, $b>0$:
$a_0 =0$; $a_1 = -a$; $b_0 =1$; $b_1 =b$
\end{itemize}
and
\begin{eqnarray}\label{k3}
\nu &=& - \frac{10 A_1^2 a^2 \alpha_4 - 36 a A_0 b \alpha_4 A_1 - 9 A_1 a b \alpha_3 + 18 b^2 \alpha_3 A_0 + 6 b^2 \alpha_2 + 36 b^2 \alpha_4 A_0^2}{k^2 a^2 b^2} \nonumber \\
\gamma &=& \frac{4 A_1^2 a^2 \alpha_4 - 12 a A_0 b \alpha_4 A_1 - 3 A_1 a b \alpha_3 + 6 b^2 \alpha_3 A_0 + 2 b^2 \alpha_2 + 12 b^2 \alpha_4 A_0^2}{k^2 a^2 b^2} \nonumber \\
w&=& \frac{1}{k a^2 b^3} [A_1 a (\alpha_4 A_1^2 a^2 - 8 b A_1 A_0 \alpha_4 a - A_1 b \alpha_3 a + 6  b^2 \alpha_3 A_0 +  b^2 \alpha_2 +  b^2 \alpha_4 A_0^2) - \nonumber \\
&&  2 A_0 (3 b^3 A_0 \alpha_3 -  b^3 \alpha_2 - 6 b^3 A_0^2 \alpha_4)] \nonumber \\
\alpha_1 &=& \frac{1}{k^2 a^2 b^3} [-4 k^2 A_1^2 a^4 A_0 b \alpha_4 + 3 k^2 A_1 a^3 b^2 \alpha_3 A_0 + 6 k^2 A_1 a^3 b^2 \alpha_4 A_0^2 + 6 b^3 A_0^2 \alpha_3 + 2 b^3 A_0 \alpha_2 + \nonumber \\
&& 12 b^3 A_0^3 \alpha_4 - \alpha_4  A_1^3 a^3 + A_1^2 b \alpha_3 a^2 - A_1 a b^2 \alpha_2 + k^2 A_1^3 a^5 \alpha_4 + 8 b A_1^2 A_0 \alpha_4 a^2 - 6 A_1 a b^2 \alpha_3 A_0 - \nonumber \\
&& 18 A_1 a b^2 \alpha_4 A_0^2 - k^2 A_1^2 a^4 b \alpha_3 + k^2 A_1 a^3 b^2 \alpha_2 - 3 k^2 a^2 A_0^2 b^3 \alpha_3 - 2 k^2 a^2 A_0 b^3 \alpha_2 - 4 k^2 a^2 A_0^3 b^3 \alpha_4] \nonumber \\
\end{eqnarray}
Thus the equation
\begin{eqnarray}\label{k4}
&& \frac{\partial u}{\partial t} - \frac{\partial^3 u}{\partial x^2 \partial t} + \frac{\partial u}{\partial x} 
\Bigg( \frac{1}{k^2 a^2 b^3} [-4 k^2 A_1^2 a^4 A_0 b \alpha_4 + 3 k^2 A_1 a^3 b^2 \alpha_3 A_0 + 6 k^2 A_1 a^3 b^2 \alpha_4 A_0^2 + 
\nonumber \\
&& 6 b^3 A_0^2 \alpha_3 + 2 b^3 A_0 \alpha_2 + 12 b^3 A_0^3 \alpha_4 - \alpha_4  A_1^3 a^3 + A_1^2 b \alpha_3 a^2 - A_1 a b^2 \alpha_2 + k^2 A_1^3 a^5 \alpha_4 + \nonumber \\
&& 8 b A_1^2 A_0 \alpha_4 a^2 - 6 A_1 a b^2 \alpha_3 A_0 - 18 A_1 a b^2 \alpha_4 A_0^2 - k^2 A_1^2 a^4 b \alpha_3 + k^2 A_1 a^3 b^2 \alpha_2 -
\nonumber \\
&& 3 k^2 a^2 A_0^2 b^3 \alpha_3 - 2 k^2 a^2 A_0 b^3 \alpha_2 - 4 k^2 a^2 A_0^3 b^3 \alpha_4] + 2 \alpha_2 u + 3 \alpha_3  u^2 + 4 \alpha_4 u^3 \Bigg) - \nonumber \\
&& \nu \frac{\partial u}{\partial x} \frac{\partial^2 u}{\partial x^2} - \gamma u \frac{\partial^3 u}{\partial x^3} =0
\nonumber \\
\end{eqnarray}
has the solutions
\begin{equation}\label{k5}
u(\xi) = \cfrac{A_0 - \det  \begin{bmatrix}
A_0 & A_1 \\
a & b
\end{bmatrix}  \exp (\xi)}{1 - b \exp(\xi)}, \hskip.25cm a>0, b<0
\end{equation}
\begin{equation}\label{k6}
u(\xi) = \cfrac{A_0 + \det  \begin{bmatrix}
A_0 & A_1 \\
a & b
\end{bmatrix}  \exp (\xi)}{1 + b \exp(\xi)}, \hskip.25cm a<0,
b>0
\end{equation}
where
\begin{eqnarray}\label{k7}
\xi = k x + \frac{t}{k a^2 b^3} [A_1 a (\alpha_4 A_1^2 a^2 - 8 b A_1 A_0 \alpha_4 a - A_1 b \alpha_3 a + 6  b^2 \alpha_3 A_0 +  b^2 \alpha_2 +  b^2 \alpha_4 A_0^2) - \nonumber \\
  2 A_0 (3 b^3 A_0 \alpha_3 -  b^3 \alpha_2 - 6 b^3 A_0^2 \alpha_4)]
+ \delta \nonumber \\
\end{eqnarray}
and $\det [\dots]$ denotes the determinant of the corresponding matrix.
\section{Concluding remarks}
For the nonlinear dynamics, chaos theory and for the
nonlinear physics in general the methods for obtaining exact analytical 
solutions of classes of nonlinear PDEs are of great interest. It seems 
however that some of these methods are more fundamental than other ones. 
In this paper we discuss the relations between  the method of simplest 
equation and the exp-function method. On the basis of our experience gained 
by application of the method of simplest equation to various nonlinear PDEs 
we consider a generalization (\ref{e9}) of the ansatz of the exp-function 
method and then we demonstrated the obtaining of exact traveling wave 
solutions of a member of the class (\ref{a1}) of nonlinear PDEs.

\begin{center}
{\sc REFERENCES}
\end{center}
$[^1]$ Frank T. D. Nonlinear Fokker - Planck equations, Springer, Berlin, 2005.\\
$[^2]$ Kantz H., D. Holstein, M. Ragwitz, N. K. Vitanov. Physica A, {\bf 342}, 2004,
       Nos. 1-2, 315 -- 321.\\
$[^3]$ Vitanov N. K., E. D. Yankulova. Chaos Solitons \& Fractals, {\bf 28}, 2006, No. 3, 
		768 -- 775 \\
$[^4]$ Panchev S., T. Spassova, N. K. Vitanov. Chaos, Solitons \& Fractals, {\bf 33},
       2007, No. 5, 1658 - 1671. \\
$[^5]$ Vitanov N. K. Proc. Roy. Soc. London A, {\bf 454} 1998, No. 1977, 2409 - 2423.\\
$[^{6}]$ Temam, R. Navier - Stokes equations: Theory and numerical analysis. AMS Chelsea
	Publishing, Providence, R. I., 2001.\\
$[^{7}]$ Vitanov N K.   Physica D, {\bf 136}, 2000, Nos. 3-4, 322 - 339. \\
$[^{8}]$ Radev S., N. Vitanov. Compt. rend. Acad. bulg. Sci., {\bf 64} 2011, No.
3, 353 - 360.\\
$[^{9}]$ Vitanov N. K. Phys. Rev. E, {\bf 62}, 2000, No. 3,  3581 - 3591.\\
$[^{10}]$ Vitanov N. K. European Physical Journal B, {\bf 15} 2000, No. 2, 349 - 355.\\
$[^{11}]$ Infeld, E., G. Rowlands. Nonlinear waves, solitons and chaos. Cambridge University
	Press, Cambridge, UK, 1990.\\
$[^{12}]$ Boeck T., N. K. Vitanov N K. Phys. Rev. E, {\bf 65} 2002, No. 3, Article number:  
	037203.\\
$[^{13}]$ Vitanov N K, I. P. Jordanov, Z. I. Dimitrova. Commun. Nonlinear Sci. Numer. 
	Simulat., {\bf 14} 2009, No. 5, 2379 - 2388.\\
$[^{14}]$ Vitanov N K, I. P. Jordanov, Z. I. Dimitrova. Applied Mathematics and Computation. 
	{\bf 215} 2009, No. 8, 2950 - 2964.\\
$[^{15}]$ Remoissenet M. Waves called solitons, Berlin, Springer, 1993.\\
$[^{16}]$ Kudryashov N A. Chaos Solitons \& Fractals, {\bf 24}, 2005, No. 5, 
	1217 - 1231. \\	
$[^{17}]$	Vitanov N. K.  Commun. Nonlinear Sci.  Numer.  Simulat. {\bf 16}, No. 3, 2011, 
	1176 - 1185.\\	
$[^{18}]$ Vitanov N. K., Z. I. Dimitrova, H. Kantz. Applied Mathematics and Computation
	{\bf 216}, 2010, No. 9, 2587 - 2595. \\	
$[^{19}]$ Kudryashov N. A. Regular \& Chaotic Dynamics, {\bf 13}, 2008, No.3,  234 - 238.\\
$[^{20}]$ Vitanov N. K. Commun. Nonlinear Sci.  Numer. Simulat. {\bf 15} 2010, 
	No. 8, 2050 - 2060.\\
$[^{21}]$ Vitanov N. K., Z. I. Dimitrova. Commun. Nonlinear Sci.  Numer. Simulat. {\bf 15}
	2010, No. 10, 2836 - 2845. \\
$[^{22}]$ Vitanov N.K., Z. I. Dimitrova, K. N. Vitanov.  Commun. Nonlinear
	Sci. Numer. Simulat. {\bf 16} 2011, No. 11, 3033 - 3044.\\
$[^{23}]$ Kudryashov N. A. Chaos Solitons \& Fractals {\bf 24}, 2005, No. 5,
	1217-1231. 
$[^{24}]$ He, J. -H., X. -H. Wu. Chaos Solitons \& Fractals, {\bf 30}, 2006, No. 3, 
	700 - 708.

\begin{flushleft}
"G. Nadjakov" Institute of Solid State Physics \\
Bulgarian Academy of Sciences \\ 
Blvd.Tzarigradsko Chausse 72\\
 1784, Sofia, Bulgaria\\
 e-mail: {\sl zdim@issp.bas.bg}
\end{flushleft}
\end{document}